# RADIO ASTRONOMY WITH MULTIBAND RECEIVERS AND FREQUENCY PHASE TRANSFER:

# Scientific Perspectives

Report from a Workshop held in Bonn on 12-14 October 2022

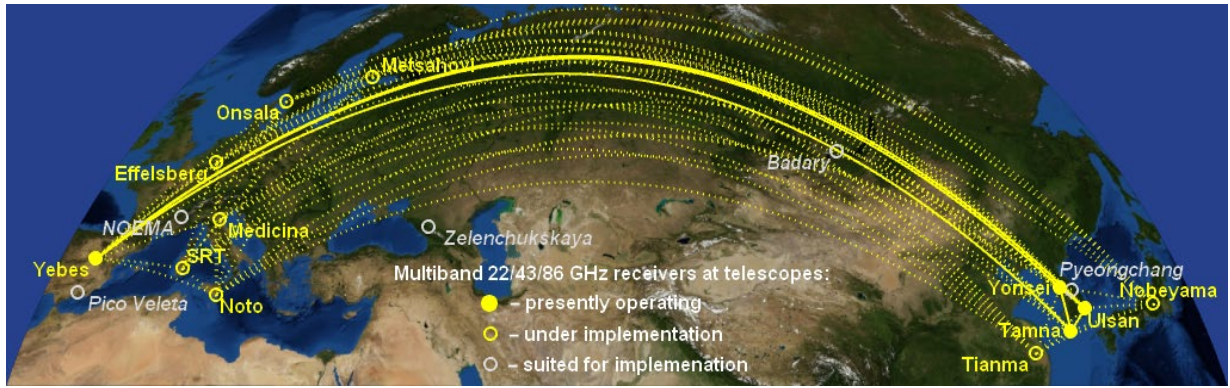

Richard Dodson[1], Cristina García-Miró[2], Marcello Giroletti[3], Taehyun Jung[4], Michael Lindqvist[5], Andrei Lobanov[6], Maria Rioja[7,1,2], Eduardo Ros[6], Tuomas Savolainen[8], Bong Won Sohn[4], Anton Zensus[6], Guang-Yao Zhao[9]

1 – ICRAR, Crawley, Australia; 2 – Yebes Observatory (IGN), Spain; 3 – IRA (INAF), Bologna, Italy; 4 – KASI, Daejeon, Korea; 5 – Onsala Space Observatory, Sweden; 6 – MPIfR, Bonn, Germany; 7 – CSIRO, Bentley, Australia; 8 – Aalto University, Espoo, Finland; 9 – IAA, Granada, Spain.

## Summary Statement

► The technique of frequency phase transfer (FPT), enabled by multiband receivers with shared optical path (SOP), is set to become a true backbone of VLBI operations at frequencies above 22 GHz. The FPT has been successfully implemented at the Korean VLBI Network (KVN), while gaining ever more prominent attention worldwide.

► Over the next few years, FPT VLBI at 22/43/86 GHz will become feasible at more than ten telescopes in Eurasia and Australia. This development would bring order of magnitude improvements of sensitivity and dynamic range of VLBI imaging at 86 GHz and deliver astrometric measurements with an accuracy of one microsecond of arc. The resulting exceptional discovery potential would strongly impact a number of scientific fields ranging from fundamental cosmology and black hole physics to stellar astrophysics and studies of transient phenomena.

► It is now the right moment for establishing a Science Working Group and a Technical Working Group for FPT VLBI in order to actively focus and coordinate the relevant activities at all stakeholder institutes and ultimately to realize the first global FPT VLBI instrument operating at 22/43/86 GHz.

# Contents

## Background

Radio astronomy and very long baseline interferometry (VLBI) undergo a period of revolutionary changes, with next generation instruments set to reach unprecedented sensitivity, spectral and angular resolution, and image fidelity. Current VLBI observations provide image sensitivity of order of 10 micro-Jansky (μJy) with the European VLBI Network (EVN) and global VLBI experiments[1], reach instrumental resolution of 20–40 microarcseconds (μas) with the Event Horizon Telescope (EHT)[2] and the Global Millimetre VLBI Array (GMVA)[3], and deliver astrometric accuracy of about 30–50 μas with the Korean VLBI Network (KVN)[4].

In the coming decade, the dynamic range of EHT imaging needs to be improved by a factor of ~50 for carrying out decisive tests for the existence of cosmic black holes[5]. A factor of ~10 improvement of dynamic range or VLBI imaging at 86 GHz would bring its effective imaging resolution on par with that of the present day EHT imaging and will allow detailed black hole studies to be made also at 86 GHz. Bringing the astrometric accuracy to below 10 μas would expand VLBI studies into a range of new scientific themes, including fundamental cosmological applications.

Success at reaching these ambitious goals is critically dependent on the capability to increase coherence time, and reduce phase noise, of VLBI measurements. At frequencies above 22 GHz, this is best achieved through the technique of frequency phase transfer (FPT)[4,6,7] which uses phases recorded at a lower frequency to calibrate phases measured at one or more higher frequencies.

Pioneered with only *nearly simultaneous* multifrequency VLBI measurements[8], the FPT technique started to produce a much stronger impact after the development of "quasi optical" multiband receivers with the shared optical path (SOP)[9,10]. Such receivers enable making *truly simultaneous* measurements at multiple frequencies, which substantially improves the quality of the phase transfer calibration.

The Korean VLBI Network (KVN) presently operates four-band (22/43/86/129 GHz) SOP receivers at three 21-meter antennas spanning across baselines of 305–478 km. The FPT operations at the KVN have led to increasing the coherence time by more than two orders of magnitude and reducing the phase noise down to ~20° at 129 GHz[11]. This has allowed the KVN to improve the detection sensitivity by more than an order of magnitude and successfully implement a source frequency phase referencing (SFPR) technique to carry out relative astrometry measurements at an exceptional ~30 μas accuracy.

The excellent scientific potential offered by the FPT and SFPR has been promptly recognized worldwide. In the USA, the FPT technique has been adopted as a key design element by the *next generation* EHT (*ng*EHT) initiative. In Europe, the Yebes 40-m telescope in Spain already operates a 22/43/86 GHz FPT capable receiver, while further compact triple-band (CTR) receivers operating at 22/43/86 GHz[12] are being implemented or planned at six more telescopes in Medicina, Noto, and SRT in Italy, Effelsberg in Germany, Onsala in Sweden, and Metsähovi in Finland. Further plans and preparations are underway for implementing multiband receivers at Sejong (Korea), Tianma (China), Nobeyama and VERA (Japan), ATCA (Australia), and NARIT (Thailand).

The eventual combination of all these telescopes with the KVN antennas is going to create the first truly global FPT VLBI network and transform the VLBI operations at 86 GHz.

In anticipation of these changes, the Bonn workshop has brought together representatives from the KVN and the seven abovementioned European telescopes, with the goals of identifying areas of the strongest scientific impact from FPT VLBI and discussing technical and organisational issues that may need to be addressed during its implementation. The outcomes of these discussions are summarized below.

## Global FPT VLBI

The main science applications of a global FPT VLBI array are going to result from the expected improvements provided by the current benchmark performance of the KVN[13,14]:

1) Increasing the coherence time at 86 and 129 GHz by more than two orders of magnitude and thereby reducing phase noise at higher frequencies to

$$\sigma_{\mathrm{ph}}(\nu) \approx 5° \; (\nu/22 \, \mathrm{GHz}) \; (\mathrm{SNR_{22GHz}}/10)^{-1/2} \; ,$$

in FPT VLBI measurements with the reference frequency of 22 GHz.

2) Performing relative astrometry measurements with the residual phase noise of

$$\sigma_{\mathrm{res}} \approx 0.005° \; (\nu/\mathrm{GHz})^{1.3} \; (\theta_{\mathrm{sep}}/1°) \; (\mathrm{SNR_{ref}}/100)^{-1/2} \; ,$$

using the SFPR with the reference frequency of 22 GHz and with a reference source detected at a signal-to-noise ratio, $\mathrm{SNR_{ref}}$.

3) Performing relative astrometry measurements with a positional accuracy of

$$\sigma_{\mathrm{pos}} \approx 50 \, \mu\mathrm{as} \; (B/500 \, \mathrm{km})^{-1} \; (\mathrm{SNR_{ref}}/100)^{-1/2} \; ,$$

which scales inversely proportional to the longest baseline, *B*, used for the measurement.

If similar performance is achieved in a global FPT VLBI array consisting of ten or more antennas (see Table 1), it should provide an **astrometric accuracy down to ~1 μas** and a **10 times higher dynamic range** of 86 GHz imaging compared to the current performance of the GMVA.

The global FPT VLBI observations at 86 GHz, referenced to 22 GHz measurement with the baseline SNR≥10, would reach an effective **image resolution of ~20 μas**, which is comparable to the image resolution of the current EHT observations (see Lu et al. 2023[15] for recent results from GMVA imaging of M87).

**Table 1. Status of existing, planned, and potential future multiband SOP receivers**

| Antenna | Receiver Band | | | | |
|---------|--------|--------|--------|---------|---------|
| | **22 GHz** | **43 GHz** | **86 GHz** | **129 GHz** | **230 GHz** |
| KVN: Yonsei | in operation | in operation | in operation | in operation | *planned* |
| KVN: Ulsan | in operation | in operation | in operation | in operation | |
| KVN: Tamna | in operation | in operation | in operation | in operation | |
| KVN: Pyeongchang | *in 2024/Q3* | *in 2024/Q3* | *in 2024/Q3* | *in 2024/Q3* | *in 2024/Q3* |
| Sejong | in operation | in operation | *possible* | | |
| Yebes | in operation | in operation | in operation | | |
| ATCA [*] | in operation | in operation | in operation | | |
| Noto | *in 2023/Q4* | *in 2023/Q4* | *in 2023/Q4* | | |
| SRT | *in 2023/Q4* | *in 2023/Q4* | *in 2023/Q4* | | |
| Medicina | *in 2024/Q2* | *in 2024/Q2* | *in 2024/Q2* | | |
| Effelsberg | *in 2024/Q2* | *in 2024/Q2* | *in 2024/Q2* | | |
| Metsähovi | *In 2026/Q1* | *In 2026/Q1* | *In 2026/Q1* | | |
| Onsala | *design* | *design* | *design* | | |
| Tianma | *planned* | *planned* | *planned* | | |
| Nobeyama | *under tests* | *under tests* | *under tests* | | |
| Mopra | *planned* | *planned* | *planned* | | |
| Pico Veleta | *possible* | *possible* | *under tests* | *possible* | *under tests* |
| NOEMA | *possible* | *possible* | *possible* | *possible* | *possible* |
| APEX | *possible* | *possible* | *possible* | *possible* | *possible* |
| Zelenchukskaya | *possible* | *possible* | *possible* | | |
| Badary | *possible* | *possible* | *possible* | | |

[*] - limited frequency range, operating in the paired-antenna mode, using single-band receivers.

## Development status of FPT VLBI

At present, FPT VLBI can be operated at four bands (22/43/86/129 GHz) with the three KVN antennas. The respective frequency bands used at the KVN are 18-26 GHz, 35-50 GHz, 80-116 GHz, 125-142 GHz[16]. The Yebes telescope has the capacity to perform FPT VLBI observations at the three lower bands (22/43/86 GHz) and combined KVN+Yebes observations are presently being carried out in the test mode. Further test and ad hoc FPT VLBI experiments are being done by combining the KVN antennas with Yebes[17] and VERA[18] antennas (with 22/43 GHz FPT receivers provided for VERA by the KVN) and the ATCA in Australia (in the paired antenna mode).

FPT VLBI operations are principally feasible with the antennas at Pico Veleta, Plateau de Bure (NOEMA), and APEX where they could be extended to frequencies up to 230 GHz and combined with the planned fourth KVN antenna in Pyeongchang. Three-band FPT VLBI operations are also feasible with the new KVAZAR antennas recently installed in Badary and Zelenchukskaya.

According to present day development planning, by the end of 2024 four antennas in Europe (Noto, SRT, Medicina, and Effelsberg) should have the capacity of performing three-band FPT VLBI observations, and several more should join in within the next few years.

## Scientific Applications of FPT VLBI

A global FPT VLBI array comprising 10+ antennas operating at 22/43/86 GHz on baselines up to 9000 km will have the most profound impact on a number of scientific areas, benefiting from the improved detection sensitivity and imaging dynamic range and the unprecedented astrometric accuracy[19]. These areas are summarized in Table 2 and in explanatory notes to it.

**Table 2. Expected main areas of scientific impact of FPT VLBI**

| Science Area | Main Goals | Impact Level |
|---|---|---|
| Cosmology | Hubble constant measurements with local galaxies at distances up to ~100 Mpc | Unique |
| Galactic dynamics | Direct proper motion and parallax measurements at distances up to ~100 kpc | Unique |
| Galactic dynamics | Most accurate determination of Solar motion inside the Galaxy and with respect to the CMB reference frame | Unique |
| Stellar astrophysics | Co-location and physics of different maser species | Unique |
| Black hole physics | Event horizon tests with magnetic field measurements | Critical |
| Black hole physics | Black hole shadow and disk-jet connection studies | Critical |
| Black hole physics | Dynamic imaging of black hole shadow and inner jet in M87 | Critical |
| Black hole physics | Hot spot motion in Sgr A* | Essential |
| GW and Transients | Localizations and followups of TDE, VHE flares and gravitational wave and neutrino detections | Essential |
| Binary SMBH | Orbital motion of radio loud companion in binary SMBH | Essential |
| AGN and jet studies | Synchrotron spectrum and rotation measure imaging | Essential |

### Cosmology and galactic dynamics

The superb astrometric accuracy expected from SFPR observations made with the three-band receivers will enable a number of unique observations to be made. Benchmarking on the KVN performance[20], a positional accuracy of $\sigma_{rel}$ = 10 µas (comparable to the expected Gaia end-of-mission parallax uncertainty for a 14[th] magnitude object[21]) can conservatively expected to be achievable with relative astrometry on ~9000 km baselines. Then, with $N_{obs}$ three-band FPT VLBI observations, it would be possible to measure:

a) yearly parallaxes for objects at distances up to $\approx 100 \text{ kpc} \sqrt{N_{obs}/6} \left(\frac{10\,\mu\text{as}}{\sigma_{rel}}\right)$ ;

b) proper motions for objects at distances up to $\approx 20 \text{ kpc} \left(\frac{v}{\text{km/s}}\right)\left(\frac{\Delta t}{\text{yr}}\right)\sqrt{N_{obs}/6} \left(\frac{10\,\mu\text{as}}{\sigma_{rel}}\right)$ ;

c) secular parallaxes for objects at distances up to $\approx 78 \text{ Mpc} \left(\frac{\Delta t}{\text{yr}}\right)\sqrt{N_{obs}/6} \left(\frac{10\,\mu\text{as}}{\sigma_{rel}}\right)$ .

In these measurements, positions of the target sources will be tracked against distant radio sources for which the magnitude of the respective effects will be below the SPFR precision. The yearly parallax and proper motion measurements will provide the most accurate distance and kinematic data throughout the Milky Way and out to the Magellanic Clouds. Measurements of proper motions and secular[22] and cosmological[23] parallaxes of radio sources in nearby galaxies will provide the most detailed account of the Solar and Galactic motion within the Local Group of galaxies[24] and with respect to the CMB rest frame. They will also yield uniquely accurate estimates of distances to a number of nearby galaxies. This would provide an independent and accurate measure of the local value of the Hubble constant[25], addressing the existing tension with its measurements performed at different redshifts.

## Stellar astrophysics

Application of SFPR to FPT VLBI data will allow for accurate cross-band image registration to be made. Already successfully tried at the KVN[26,27], these efforts will allow accurately co-locating different maser species (H20 at 22 GHz, SiO at 43 and 86 GHz) in order to obtain a unique handle on the physical conditions in a variety of stellar maser sources.

The robust cross-band registration will also be used for carrying out the most detailed studies of morphological and spectral evolution of contact binaries with strong continuum radio emission, including the objects like RS Oph, AE Aqr, and AR Sco. Applied to colder M-type stars, it will also allow for robust tracking of star spot evolution and accurate measurements of their rotational periods to be made.

## Black hole physics

The combination of improved sensitivity and accurate cross-band image registration will be essential for making robust estimates of magnetic field strength near supermassive black holes, based on the core shift measurements, turnover frequency/flux density imaging and rotation measure mapping. These measurements will provide first reliable probes of the magnetic field on scales down to several hundreds of gravitational radii, where such probes could yield decisive tests for discerning between the canonical black holes and a broad class of their horizonless alternatives[28].

FPT VLBI images reaching a dynamic range of $D_{86GHz}$ at 86 GHz will have an effective image resolution of $\approx 100 \, \mu as \, / \sqrt{D_{86GHz}}$. For many objects, including M87 and Sgr A*, this resolution will match the current resolution of the EHT imaging at 230 GHz. This will allow for studying photon ring and disk-jet connection (stronger manifested at 86 GHz) for M87 and a number of galaxies which may be too weak for making robust images with the EHT.

FPT VLBI at 86 GHz will also provide an excellent tool for studying dynamic evolution of the material in the accretion disks in M87 and Sgr A*. Movies of disk-jet evolution can be made for M87 on timescales of two weeks. At 86 GHz, the hot spot evolution in Sgr A* can be traced at an $N_\sigma$ level, while beating the scattering, with observations reaching the image signal-to-noise ratio of $\approx 40 \, N_\sigma \left(\frac{\lambda}{cm}\right) \left(\frac{B_{max}}{1000 \, km}\right)^{-1}$.

## Transient phenomena

The excellent multifrequency and astrometric capabilities of FPT VLBI will allow it to be used as an efficient tool for localization and dedicated followups of a whole range of transient phenomena, from tidal disruption events, gamma-ray bursts, high energy flares, and supernova explosions, to gravitational wave and neutrino detections. Accurate multifrequency measurements will be particularly important for accurately assessing the spectral and polarization properties of the radio counterparts of each of these transient phenomena.

### Binary SMBH

Relative astrometry observations with FPT VLBI will allow tracing fine orbital motions in supermassive binary black holes with sub-parsec separations between the companions. If at least one of the companions remains radio loud and its position can be measured at an astrometric accuracy of $\sigma_{rel}$, then its orbital motion can be detected over a time span of the half of the orbital period, $T_{orb}$, at a significance level of

$$N_\sigma \approx \left( \frac{10\ \mu as}{\sigma_{rel}} \right) \sqrt{\frac{N_{obs}}{6}} \left( \frac{T_{orb}}{10\ yr} \right)^{\frac{2}{3}} \left( \frac{M_{bh}}{10^9\ M_\odot} \right)^{\frac{1}{3}} \left( \frac{D_A}{1\ Gpc} \right)^{-1} .$$

This will allow exploring a large sample of radio loud AGN in which presence of binary supermassive black holes at sub-parsec separation can be suggested from indirect observational evidence.

Similar studies can be made for detecting fine scale changes in gravitational lens systems, allowing for refined modelling of the lensing object and its environment, including the putative dark matter halos.

### Compact relativistic objects and relativistic outflows

Multifrequency and astrometric capabilities of FPT VLBI will provide an excellent set of tools for dedicated studies of compact relativistic objects, including radio loud AGN and XRB sources, and detailed investigations of physical properties of collimated relativistic outflows. They will allow using brightness temperature, turnover frequency/flux density, and Faraday rotation as effective tools for probing physical conditions near the AGN/XRB and inside the outflows. Kinematic and opacity (core shift) properties of the outflows will be investigated in detail using the relative astrometric measurements and bona fide image alignment provided by FPT VLBI.

All these studies could be done both on an object-to-object basis and as part of larger scale survey type efforts.

## Technical implementation of FPT VLBI

A quick review of the current technical developments at different institutes and observatories working on SOP receivers and three-band FPT VLBI has been undertaken at the workshop and combined with the summary information from a short technical meeting on the triple-band SOP receivers organized in late August within the framework of the EVN TOG activities. This review indicates that the individual institutional efforts remain rather heterogeneous and, in some instances, resulting in differences and variability of technical specifications and modes of operation of the receivers.

In view of this observation, a dedicated effort may be needed for compilation and assessment of the technical capabilities of all existing triple-band receivers and receivers which are presently under design and construction. This includes, in particular, the band specifications (e.g. minimally accepted frequency ranges in each of the three operational bands) and polarization properties of the receivers.

It would be highly desirable if a common design framework could be defined for the receivers, at least in those aspects of the design which are not subject to the necessary customization (e.g., the optical path and dichroic filter specifications which are strongly influenced by the local conditions such as the dimensions and existing load of the focal cabins at individual telescopes).

The technical implementation would also benefit strongly from defining a common approach toward the recording and backend equipment to be used and further developed for the FPT VLBI operations. This could be facilitated through stronger involvement of the EVN TOG and the GMVA Technical Group and perhaps also aided by further capacities and opportunities that may exist within the recently established Global VLBI Alliance.

## Organizational implementation of FPT VLBI

Until now, all FPT VLBI observations have been organized and carried out either within the KVN and EAVN or by means of ad hoc experiments combining the KVN antennas with external telescopes (Yebes, Nobeyama, VERA antennas, ATCA).

With the expected gradually increasing number of FPT-capable telescopes, it would eventually become necessary to provide an organizational framework for carrying out FPT VLBI observations on a more regular basis. The provisions provided presently by the EVN, Global VLBI, and the GMVA can in principle be suitable for this purpose, although a larger flexibility for time allocation (going away from the rigid session-type structuring of VLBI observations) would be preferred for most of the key science areas for FPT VLBI and in particular for studies relying on multi epoch (e.g. yearly parallax or dense imaging projects) and rapid response capabilities.

Engagement in this type of studies is also likely to result in increased demands for a continued and dedicated access to the correlator facility and the availability of advanced pipelining facility. At present, these needs can be satisfied by the JIVE correlator in Dwingeloo and the GMVA correlator in Bonn, while the correlator facilities at the KVN and the IRA Bologna would have to be upgraded to satisfy the anticipated needs of global FPT VLBI.

## Specific requirements arising from the scientific priorities of FPT VLBI

The scientific focus of FPT VLBI described above results in several specific and/or new technical and operational requirements which should be discussed and addressed within the next two years.

The anticipated increased recording rate (due to simultaneous three-band VLBI operations) must be accommodated from the technical, logistical, and organizational perspectives. This accommodation may demand closer coordination between the EVN, GMVA, and EAVN operations and structures.

Increased time commitments and/or dedicated time allocations for FPT VLBI are likely to be needed on the part of all participating telescopes and institutes.

FPT VLBI should be potentially considered as the future standard mode of VLBI operations at 22/43/86 GHz. The transition to this state and its relation and connection to the present day GMVA, EVN, and EAVN operations should to be further elaborated and formulated.

Several scientific areas, including high dynamic range imaging and observations of Sgr A* would benefit strongly from closing the existing gap between the European antennas and the KVN and enabling FPT VLBI on Australian telescopes. Eventual inclusion of ATCA, Mopra, Tianma would be a logical first step in this direction.

Enabling FPT VLBI on one or more Northern hemisphere antennas at Eastern longitudes between 40° and 100° (such as the KVAZAR antennas in Zelenchukskaya and Badary and the planned large millimetre wavelength telescope in Qitai, near Urumqi) would be the ultimate step for closing the Europe-KVN gap.

On a longer term, potential synergies with the ngVLA would inevitably become an important factor. Early engagement with the NRAO and ngVLA community would be beneficial in this regard.

## Conclusions from the Workshop

Considering the information and arguments presented and discussed at the workshop, the following summarizing conclusions and suggestions can be formulated:

1. The gradually increasing number of telescopes equipped with triple-band (22/43/86 GHz) receivers presents ample scientific opportunities arising from implementation of the FPT and SFPR techniques for VLBI. It is important to actively engage in this emerging area of VLBI research.

2. While the final modalities of global FPT VLBI operations would have to be elaborated and formulated in the course of a broader discussion, several measures and steps should be taken already now in order to streamline and shape up the ongoing developments.

3. Information exchange on all technical aspects of implementation of SOP receivers for FPT VLBI should be intensified and channelled through joint activities of the EVN TOG and the GMVA Technical Group. Ideally, a dedicated Technical Working Group for FPT VLBI should be formed, whose main task would be to coordinate and focus the activities at the individual observatories and institutes.

4. It would be very beneficial to begin carrying out regular tests of the FPT VLBI, initially with the participation of the KVN antennas and Yebes telescope while other telescopes would gradually join this activity. This could be one of the tasks for the Technical Working Group.

5. A *Science Working Group* should be formed in order to explore the scientific potentials of FPT VLBI in full breadth and depth and to engage in a continued interaction with the *Technical Working Group* in order to provide the best match between the scientific goals and technical capabilities of FPT VLBI.

6. The immediate goals for the Science Working Group and the Technical Working Group should be the development of a comprehensive *White Paper on Key Science with Multiband Receivers* and a robust *Technological Roadmap toward a Global FPT VLBI Instrument*.

7. Further discussions of broader topics such as extending FPT VLBI capabilities to other frequencies and exploiting synergies with other instruments and facilities should be initiated and supported, using all suitable channels and platforms, including the Global VLBI Alliance.